\documentclass[letterpaper,conference]{IEEEtran}
\usepackage{amssymb}
\usepackage{verbatim}
\usepackage{graphicx,epsfig}
\usepackage{amsfonts}
\usepackage{subfigure}
\usepackage[normalem]{ulem}
\usepackage{amsmath}
\usepackage{amsthm}
\usepackage[table]{xcolor}

\begin{document}

\title{Caching Piggyback Information for Efficient Index Code Transmission}

\author{Jalaluddin Qureshi\\
Department of Electrical Engineering\\
Namal College, Mianwali, Pakistan}

\maketitle

\begin{abstract}
The index coding problem is a fundamental transmission problem arising in content distribution and wireless networks. Traditional approach to solve this problem is to find heuristic/ approximation minimum clique partition solution on an appropriately mapped graph of the index coding problem. In this paper we study index code for unicast data flow for which we propose updated clique index coding (UCIC) scheme, UCIC piggybacks additional information in the coded symbol such that an unsatisfied client can update its cache. We show that UCIC has higher coding gain than previously proposed index coding schemes, and it is optimal for those instances where index code of minimum length is known.
\end{abstract}

\textbf{\textit{Keywords:}} \textit{Network Coding; Erasure Correction; Graph Theoretic Algorithms; Content Distribution; Transmission Minimization;}

\newtheorem{axiom}{Axiom}
\newtheorem{lemma}{Lemma}
\newtheorem{theorem}{Theorem}
\newtheorem{observation}{Observation}
\newtheorem{definition}{Definition}

\section{Introduction} \label{sect:Introduction}
The \emph{index coding problem}, a computationally intractable problem, is an instance of data transmission problem to a set of $n$ clients $\mathcal{C} = \{c_1, c_2, ..., c_n\}$ by a server having a set of $k$ input symbols $\mathcal{P} = \{p_1, p_2, ..., p_k\}$, $p_j\in \mathsf{F}_2^\mathsf{b}$, given the \emph{side information} knowledge, known as the \textit{has set} $H_i\subset \mathcal{P}$, the server has about the set of symbols each client $c_i\in \mathcal{C}$ has in its cache, and the set of symbols each client wants, known as the \textit{want set} $W_i\subseteq \mathcal{P} \backslash H_i$, such that the total number of transmissions are minimized. The code constructed to satisfy the index coding problem are known as \emph{index code} and is denoted by the set $\mathcal{D}$ whose cardinality is given by $\ell=|\mathcal{D}|$. This problem formulation finds application in several transmission problems using broadcast channel, most notably content distribution network~\cite{Borst10, Ali13}, wireless network~\cite{Ong12, Katti06} and network coding~\cite{Rouayheb10}.

In a content distribution network such as VoD (Video-on-Demand) libraries used by YouTube, servers close to the users at network edge (which we call distribution server) cache popular videos during off-peak hours from central server, when communication systems are under-utilized, so that network congestion and performance bottleneck can be mitigated during peak traffic hour. Such transmission architecture takes advantage of decrease in memory cost at a rate higher than that of transmission gear. It is expected that the client's request during peak hours can be served by distribution server, however this may not necessarily always be the case, as the client may request a video not available at the distribution server, in which case a request has to be placed to the central server. Over a given time interval, the central server receive several such requests. For such scenario the central server can take advantage of the cached information at each of the distribution server to transmit index code.

\begin{figure}
\begin{center}
\includegraphics[width = 0.5\textwidth]{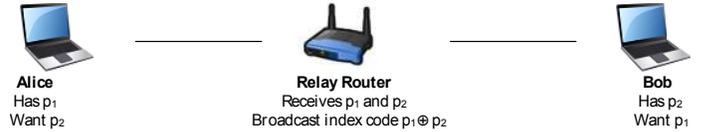}
\end{center}
\caption{Alice and Bob want to exchange symbols through a wireless relay router. After receiving $p_1$ and $p_2$ from Alice and Bob respectively, instead of forwarding $p_1$ and $p_2$ in two different time slots, the relay router can broadcast the index code $p_1\oplus p_2$ in one time slot which will satisfy the want set of both Alice and Bob, and save one transmission by taking advantage of the cached symbols in the has sets of Alice and Bob.} \label{fig:alice_bob} 
\end{figure}

Similarly such transmission problem also arises in wireless networks. Due to the broadcast nature of the wireless transmission, a transmitted symbol is overheard by several neighboring stations in addition to the intended station. These overheard symbols can be treated as cached symbols, and the relay router can use the knowledge of side information of these cached symbols to construct index code to satisfy the symbols requested by each of the neighboring stations. Katti \textit{et al.} showed the application of index code (which they call opportunistic coding in their work) for routing in wireless mesh network and demonstrate its throughput gain over non-index code based routing scheme on wireless test-bed~\cite{Katti06}. In a time division multiple access (TDMA) scheme such as satellite communication, each ground terminal transmit data to the satellite on the uplink, and the satellite then retransmit the received data on the downlink~\cite{Sklar01}. Ong and Ho showed that instead of retransmitting the data, the satellite can construct index code, taking advantage of the cached information each ground terminal has (i.e. the data it wants to transmit) so that the data transmitted on the downlink is minimized~\cite{Ong12}. A simple illustrating example of the application of index code for wireless relay network is shown in Figure~\ref{fig:alice_bob}.

It has also been shown that any instance of the network coding problem can be reduced to the index coding problem, and that these two problems are equivalent in the linear case~\cite[Theorem 5]{Rouayheb10}. Network code finds applications in several transmission networks such as multi-source content distribution, routing protocols, distributed storage and error correction. Therefore an efficient solution to the index coding problem can be used to construct efficient network codes.

Despite the common application of index code in transmission networks, finding the length of index code corresponding to the minimum length $\ell^*$, such that $\ell^*=min\{\ell\}=|\mathcal{D}^*|$, is a NP-hard problem~\cite[Theorem 5]{Ziv06}. Optimal index code $\mathcal{D}^*$ can only be found for few classes of the index coding problem, namely, single-uniprior side information~\cite{Ong12}, near extreme rates~\cite{Dau14} and multicast transmission~\cite{Plank05}.

The computational intractability of this problem calls for efficient coding techniques to solve the index coding problem. In this paper we propose a drastically different approach to construct index code which we call updated clique index coding (UCIC). UCIC piggybacks additional information in its coded symbol, so that some of the requesting clients unable to satisfy their request from the transmission of a coded symbol are able to store additional symbols in their cache. Despite its algorithmic simplicity, UCIC has higher coding gain than all the previously proposed index coding schemes, and UCIC constructs $\mathcal{D}^*$ for those classes of unicast index coding problems where $\mathcal{D}^*$ is known.

The rest of the paper is organized as follow. We first present existing problem formulation in Section~\ref{section:formulation}, and use this formulation to discuss previously proposed algorithms in Section~\ref{section:related}. A description of UCIC algorithm and its computational complexity along with a motivating example is given in Section~\ref{section:proposed}. We then present simulation results in Section~\ref{section:simulation} and give proof of its optimal performance for single-uniprior side information index coding problem in Section~\ref{section:optimal}. We then conclude with discussion on potential future work in Section~\ref{section:future} and conclusion in Section~\ref{section:conclusion}.

Since this paper extensively uses concepts from graph theory, definition of these terminologies and graph optimization problems are given in the Appendix at the end of the paper.

\section{Problem Formulation}~\label{section:formulation}

For the index coding problem defined in Section~\ref{sect:Introduction} we assume that the unicast data flow is represented by a singleton has set, i.e. $k=n$, this as shown by Lemma~\ref{lemma:one_pk} does not affect the result for the general unicast transmission data flow.

\begin{lemma}\label{lemma:one_pk}
For the unicast index coding problem, without loss of generality we only need to consider the case of a single unique symbol being requested by each client (single-unicast)~\cite[Lemma 1]{Birk06}.
\end{lemma}

A client is defined as a \textit{satisfied client} if $W_i=\emptyset$, and an \textit{unsatisfied client} otherwise. \textit{Coding gain} is a ratio defined as $\frac{k}{\ell}$. The symbol has probability $p_{has}$ is defined as the uniform expected probability that client $c_i$ has symbol $p_j$, $j\neq i$. In this paper, we study scalar-linear index coding for unicast data flow. A linear code $\mathcal{D}$ is defined as a linear subspace of the vector space $\mathsf{F}_\mathsf{q}^\mathsf{b}$, and non-linear code otherwise. In scalar coding, the symbol can not be split in to smaller symbols. To solve the index coding problem the following problem formulations have been proposed to construct index code.

\textit{Side-information digraph $G=(V,E)$}~\cite{Ziv06,Dau14}. When constructing a side-information graph, single-unicast data flow is assumed, i.e. unique singleton want set (see Lemma~\ref{lemma:one_pk}). In $G$ vertex $v_i\in V$ corresponds to symbol $p_i$ requested by client $c_i$. And there exist a directed edge (arc) from vertex $v_i$ to vertex $v_j$ if client $c_i$ has symbol $p_j$. The side information graph is also known as dependency graph in~\cite{Chaudhry11}.

A matrix $\mathcal{A}=a_{ij}\in \mathsf{F}_2^{k\times k}$, is said to fit the side information graph $G$ if $a_{ii}=1$, $a_{ij}=0$ whenever $i\neq j$ and $(i,j)\notin E$, and $a_{ij}=\mathsf{a}\in \{0, 1\}$, i.e. free entries, whenever $i\neq j$ and $(i,j)\in E$.

\textit{Instantly Decodable Coding (IDC) graph $K$}~\cite{Chaudhry08, Wang10}. In $K$ a vertex $v_i$ corresponds to a symbol $p_i$ requested by client $c_i$, and an edge between vertices $v_i$ and $v_j$ exist if $p_i\subseteq H_j$ and $p_j\subseteq H_i$. This implies that, if there exist an edge between $v_i$ and $v_j$, then the coded symbol $p_i\oplus p_j$ can be instantly decoded by clients $c_i$ and $c_j$. For index coding problem with multicast data flow, there also exist an edge between two vertices in $K$ if both the clients want the same symbol. In this paper however we only study unicast data flow.

A heuristic minimum clique partition algorithm on $K$ to solve the index coding problem is based on the following idea. Each clique in $K$ corresponds to a coded symbol which satisfies the transmission request of some clients. Therefore by partitioning $K$ into minimum number of disjoint cliques the length of index code $\ell$ can be minimized.

\textit{Information-flow digraph $I$}~\cite{Ong12}. In an information-flow graph, it is assumed that each of the clients has singleton unique symbol as side information, with no restriction on the set of symbols it wants. A vertex $v_i$ in $I$ represents that client $c_i$ has symbol $p_i$. There exist a directed edge from vertex $v_i$ to vertex $v_j$ if client $c_j$ want symbol $p_i$.

\textit{Bipartite index coding digraph $B$}~\cite{Tehrani12}. In $B$, there exist a vertex for every $c_i\in \mathcal{C}$ and $p_j\in \mathcal{P}$, and a directed edge from $v_i$ to $v_j$ if $c_i$ has symbol $p_j$, and a directed edge from $v_j$ to $v_i$ if client $c_i$ has requested symbol $p_j$. The solution for bipartite index coding is limited for single-unicast and single-multicast transmission.

\begin{figure*}
\begin{center}
\includegraphics[width = 0.75\textwidth]{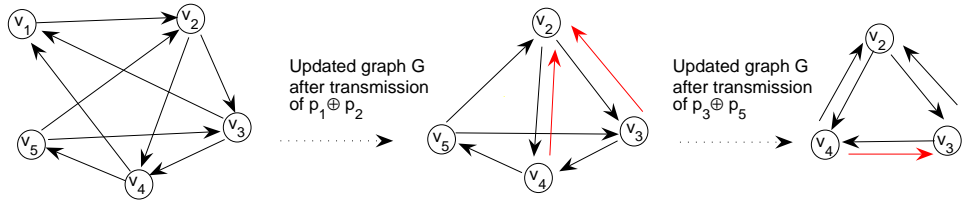}
\end{center}
\caption{Illustration of the updated graph $G$ at each iteration for the motivating example given in Section~\ref{section:motivating}.} \label{fig:graph_G}
\end{figure*}

\section{Related Work}\label{section:related}

\begin{theorem}\label{th:minrk2}
The minimum length $\ell^*$ of a linear index code for a side information graph $G$ equals $\mathsf{minrk_2}(G)$, of matrix $\mathcal{A}$ which fits graph $G$~\cite[Theorem 5]{Ziv06}.
\end{theorem}

\noindent The value of $\mathsf{minrk_2}(G)$ is bounded as,

 \[
\omega(\overline{G})\leq \mathsf{minrk_2}(G)= \ell^*\leq\varphi(G),
\]

\noindent where $\omega(\overline{G})$ is the clique number in $\overline{G}$, and $\varphi(G)$ is the minimum clique partition size of $G$. The problem $\mathsf{minrk_2}(G)$, find the minimum rank of matrix $\mathcal{A}$ which fits graph $G$, and is a NP-hard problem.

Given the NP-hardness in constructing $\mathcal{D}^*$, Birk and Kol in their original index coding paper~\cite{Birk98} and in their successive work~\cite{Birk06} proposed a heuristic minimum clique partition algorithm which they called Least Difference Greedy (LDG) algorithm to construct $\mathcal{D}$. LDG constructs a suboptimal minimum clique partition solution on the IDC graph $K$. Chaudhry and Sprinston~\cite{Chaudhry08} expanded on the work of Birk and Kol, and evaluated the performance of a $\frac{2}{3}$-approximation minimum clique partition algorithm called color saving algorithm adopted from the work of Hassin and Lahav~\cite{Hassin94} with a divide-and-conquer based greedy heuristic called sparsest set clustering algorithm. In this work Chaudhry and Sprinston showed higher coding gain of color saving algorithm over sparsest set clustering algorithm. 

The algorithmic approach proposed by Birk and Kol continues to be adopted in other generalizations of the index coding problems with unicast and multicast data flow. Shum \textit{et al.} proposed a heuristic minimum clique partition algorithm (called Partition) to solve the index coding problem on broadcast wireless erasure channel, assuming using coded symbols as side information, and a relay node as a helper~\cite{Shum12}. Wang \textit{et al.} used a heuristic minimum weighted clique partition algorithm to minimize the transmission time to multicast data to clients in a multi-rate wireless network~\cite{Wang10}. Dong \textit{et al.} used a heuristic minimum vertex coloring algorithm (called sequential coloring algorithm) as part of a dissemination protocol in wireless sensor network to efficiently retransmit erased packets~\cite{Dong11}. The minimum vertex coloring problem on the complementary graph $\chi(\overline{G})$, is equivalent to the minimum clique partition problem on the original graph $\varphi(G)$ (both of which are NP-hard problems).

Given the application of the index code in content distribution and wireless networks, and the wide adaptation of heuristic minimum clique partition algorithm to solve the index coding problem, it remains important to question whether there exist other solutions which improve on the performance of this de facto algorithmic approach to construct index codes?

In this paper we propose updated clique index coding (UCIC) scheme to improve on the performance of existing solutions. UCIC piggybacks additional information (called piggyback symbol) in the coded symbol so that the cache of unsatisfied clients can be incremented. The index code generated by UCIC is coded using the computationally simple XOR addition, and can similarly be decoded using XOR addition, making the encoding and decoding operation computationally simple. We show that UCIC has higher coding gain than previously proposed index coding schemes.

We further show that UCIC construct $\mathcal{D}^*$ for those classes of index coding problems where $\mathcal{D}^*$ is known. In our list of future work, we demonstrate the potential performance gain of UCIC over complementary index coding~\cite{Chaudhry11} and bipartite index coding~\cite{Tehrani12} solutions, which have been recently proposed as efficient solutions for the single-multicast and multiple-multicast data flow, in addition to the unicast data flow. In a single-multicast data transmission problem, all clients have non-unique singleton want set such that there exist at least two clients with different want sets. A multiple-multicast data transmission problem is characterized as $|W_i|<k, \forall i$, and $(W_i\cap W_j) \subseteq W_i, i\neq j$.

\section{Proposed Algorithm}\label{section:proposed}

\subsection{Motivating Example}\label{section:motivating}
We illustrate the coding gain of UCIC with the aid of a simple motivating example. Consider the following index coding problem given as, $W_i=\{p_i\}$, $H_1=\{p_2\}$, $H_2=\{p_3, p_4\}$, $H_3=\{p_1, p_4\}$, $H_4=\{p_1, p_5\}$, and $H_5=\{p_2, p_3\}$. Using the traditional approach of solving index coding problem by mapping this problem to graph $K$ and then running a heuristic minimum clique partition algorithm on the graph $K$~\cite{Birk98, Birk06, Chaudhry08}, it is easy to verify that the resulting graph $K$ has no edge, in which case the LDG and color-saving algorithms like any other minimum clique partition algorithms, including the exhaustive search algorithm, will partition the graph in to five cliques, resulting in index code of length five.

For our proposed algorithm, the server chooses clique of smallest size in $K$ whose vertices are given by the set $\mathcal{Y}_b$, $\mathcal{Y}_b\subset \mathcal{P}$. The algorithm then searches for $p_i \not\in\mathcal{Y}_b$ such that $W_i\neq \emptyset$ and $p_i\subseteq H_j, \forall j : p_j\in \mathcal{Y}_b$, i.e. $p_i$ should be in the want set of $c_i$, and $p_i$ should be in the has set of all clients which are elements of $\mathcal{Y}_b$. A piggyback coding symbol $p_i$ satisfying these conditions and incrementing the has set of some unsatisfied clients is then selected.

For the given example, UCIC chooses $\mathcal{Y}_b=\{p_1\}$, and selects $p_2$ as the piggyback symbol. The transmitted symbol $p_1\oplus p_2$ only satisfies the request of $c_1$, and allows $c_3$ and $c_4$ to update their has set as, $H_3=\{p_2, p_4\}$, $H_4=\{p_2, p_5\}$ (without loss of ambiguity $p_1$ has now been discarded from $H_3$ and $H_4$). The symbol $\oplus$ denotes the XOR addition operation. Since $c_1$'s request has now been satisfied, UCIC prunes $p_1$ from the graph $G$ by removing vertex $v_1$ and arc $(v_1, v_2)$, and adding two arcs $(v_3, v_2)$ and $(v_4, v_2)$ in $G$. The resulting graph $K$ now has two edges $(v_2, v_3)$ and $(v_2, v_4)$.

The smallest clique in $K$ is now given as $\mathcal{Y}_b=\{p_5\}$, and symbol $p_3$ can be used as a piggyback symbol. The transmitted coded symbol $p_3\oplus p_5$ satisfies the request of $c_5$ and increment the has set of $c_4$ as $H_4=\{p_2, p_3\}$. UCIC then prunes $p_5$ from the $G$, and adds arc $(v_4, v_2)$ in $G$. The resulting graph $K$ is now a complete graph, and UCIC then transmits $p_2\oplus p_3\oplus p_4$, satisfying the request of $c_2, c_3$ and $c_4$. An illustration of the updated graph $G$ at each step of the algorithm iteration is shown in Figure~\ref{fig:graph_G}.

The coding gain of UCIC for this example is given as $\frac{5}{3}=1.67$, while that of LDG and color-saving is given as $1$. The resulting index code $\mathcal{D^*}=\{p_1\oplus p_2, p_3\oplus p_5, p_2\oplus p_3\oplus p_4\}$ has length of three symbols. The $\mathsf{minrk_2}(G)$ for this instance of index coding problem is equal to three (found using exhaustive search), as shown below,

\begin{align*}
\mathcal{A}=
\begin{pmatrix}
1 & \mathsf{a} & 0 & 0 & 0\\
0 & 1 & \mathsf{a} & \mathsf{a} & 0\\
\mathsf{a} & 0 & 1 & \mathsf{a} & 0\\
\mathsf{a} & 0 & 0 & 1 & \mathsf{a}\\
0 & \mathsf{a} & \mathsf{a} & 0 & 1\\
\end{pmatrix}
\longrightarrow
\begin{pmatrix}
1 & 1 & 0 & 0 & 0\\
0 & 1 & 0 & 1 & 0\\
0 & 0 & 1 & 0 & 1\\
1 & 0 & 0 & 1 & 0\\
0 & 0 & 1 & 0 & 1\\
\end{pmatrix}.
\end{align*}

\noindent Therefore our proposed solution is optimal for this instance of index coding problem.

\begin{table}
\caption{UCIC Pseudocode.}
\label{table:ucic_pseudocode}
\begin{tabular}{|l|}
\hline
\textbf{Input} - Side information graph $G=(V,E)$.\\
\textbf{Output} - Index code $\mathcal{D}$.\\
\hline
\\
\textbf{While} $V\neq\emptyset$\\
\\
\ \ \textbf{Step 1}\\
\ \ \ \ Generate graph $K$ from graph $G$.\\
\ \ \ \ Heuristic minimum clique partition algorithm on $K$ or\\
\ \ \ \ Heuristic minimum vertex coloring algorithm on $\overline{K}$\\
\\
\ \ \ \ \textbf{Outputs}: Suboptimal minimum clique partition, given as,\\
\ \ \ \ $\mathcal{Q}=\{\mathcal{Y}_1, \ldots \mathcal{Y}_r$\},\\
\ \ \ \ $r$ is the number of disjoint cliques and $u\in\{1,\ldots,r\}$.\\
\\
\ \ \textbf{Step 2}\\
\ \ \ \ Group all the cliques of minimum length in set $\beta$,\\
\ \ \ \ $\beta\leftarrow\{\mathcal{Y}_b : \mathsf{min}(|\mathcal{Y}_u|)=\mathcal{Y}_b\}$.\\
\\
\ \ \textbf{Step 3}\\
\ \ \ \ Search for the piggyback coding symbol (pbs) which increments the\\
\ \ \ \ has set for maximum number of unsatisfied clients when coded with\\
\ \ \ \ symbols which are elements of $\mathcal{Y}_b$.\\
\\
\ \ \ \ \ \ $\mathsf{pbs\_found}$ = \textbf{false}\\
\ \ \ \ \ \ \textbf{For} $\forall \mathcal{Y}_b : \mathcal{Y}_b\in \beta$\\
\\
\ \ \ \ \ \ \ \ \ \textsf{GreedySearch}($\mathcal{Y}_b$)\\
\\
\ \ \textbf{Step 4a}\\
\ \ \ \ Code the piggyback symbol and transmit the coded symbol.\\
\ \ \ \ \textbf{If} $\mathsf{pbs\_found}$ is \textbf{true}\\
\ \ \ \ \ \ \ XOR the piggyback symbol and all symbols elements of $\mathcal{Y}_b$.\\
\ \ \ \ \ \ \ Transmit the coded symbol.\\
\\
\ \ \ \ \textbf{Step 4b}\\
\ \ \ \ \ \ \ Update graph $G$.\\
\ \ \ \ \ \ \ $V\leftarrow\{V\backslash v_j, \forall v_j : v_j\in \mathcal{Y}_b\}$, \\
\ \ \ \ \ \ \ Add an arc from $c_m$ to $c_i$ in $G$ if $p_i$ has been added to $H_m$.\\
\\
\ \ \textbf{Step 5}\\
\ \ \ \ \textbf{Else If} $\mathsf{pbs\_found}$ is \textbf{false}\\
\ \ \ \ Use the traditional solution, as piggyback symbol cannot be found,\\
\ \ \ \ and exit the while loop.\\
\ \ \ \ \ \ \ \textbf{For} $\forall \mathcal{Y}_u : \mathcal{Y}_u\in\mathcal{Q}$\\
\ \ \ \ \ \ \ \ \ \ XOR all the symbols which are elements of $\mathcal{Y}_u$.\\
\ \ \ \ \ \ \ \ \ \ Transmit the coded symbol.\\
\ \ \ \ \ \ \ \textbf{Break}\\
\hline
\end{tabular}
\end{table}

\begin{table}
\caption{GreedySearch($\mathcal{Y}_b$) Function.}
\label{table:greedysearch}
\begin{tabular}{|l|}
\hline
\textbf{Input} - $\mathcal{Y}_b$.\\
\textbf{Memory} - Pair of piggyback symbol $p_i$ and clique $\mathcal{Y}_b$ for which\\
maximum number of clients increment their has sets, if a piggyback\\
symbol exist.\\
\hline
\\
\textbf{For} $\forall v_i\in G : v_i\not\in\mathcal{Y}_b$\\
\\
\ \ \%If $p_i$ is in the has set of all clients members of the clique $\mathcal{Y}_b$.\\
\ \ \textbf{If} $p_i\subseteq H_j, \forall c_j : v_j\in \mathcal{Y}_b$\\
\\
\ \ \ \ \textbf{For} $\forall v_m\in G : v_m\not\in\mathcal{Y}_b, m\neq i$\\
\\
\ \ \ \ \ \ \%If client $c_m$ has all symbols which are elements of the clique $\mathcal{Y}_b$,\\
\ \ \ \ \ \ \%but does not have the piggyback symbol $p_i$.\\
\ \ \ \ \ \ \textbf{If} ($p_j\in H_m, \forall p_j : v_j\in \mathcal{Y}_b$) \textbf{and} ($p_i\notin H_m$)\\
\ \ \ \ \ \ \ \ $\mathsf{pbs\_found}$ = \textbf{true}\\
\ \ \ \ \ \ \ \ Increment (number of unsatisfied clients increasing their has set\\
\ \ \ \ \ \ \ \ from the transmission of $p_i$ and $p_j : v_j\in \mathcal{Y}_b$ coded together.)\\
\hline
\end{tabular}
\end{table}

\subsection{Pseudocode}
A pseudocode of UCIC is given in Table~\ref{table:ucic_pseudocode} and~\ref{table:greedysearch}. UCIC runs on top of heuristic/ approximation minimum clique partition algorithm, alternatively it can also run on top of a greedy algorithm which searches for cliques of size equal to one or two. The information about cliques of minimum length is then fed into the set $\beta$ (Step 2 of the algorithm).

Step 3 of the algorithm is the main contribution of UCIC. For all cliques of minimum length $\mathcal{Y}_b\in \beta$, the GreedySearch function searches for piggyback symbol which will maximize the number of unsatisfied clients whose has set can be incremented. GreedySearch stores the number of unsatisfied clients incrementing their has set from multiple pairs of $(\mathcal{Y}_b, p_i)$ and chooses the pair which maximizes the number of clients incrementing their has set from the transmission of piggyback symbol $p_i$ coded with $p_j, \forall p_j$, such that $v_j\in \mathcal{Y}_b$.

After coding $p_i$ with $p_j$, the coded symbol is transmitted (Step 4a). Vertices of satisfied clients are removed from graph $G$, and arcs added in $G$ to reflect increment in the has set of unsatisfied clients (Step 4b). The algorithm then repeats until either the want set of all clients are satisfied or the algorithm is unable to find a piggyback symbol. If the UCIC algorithm is unable to find a piggyback symbol, it then resorts to the traditional method of generating coded symbols by coding all symbols in the same clique (Step 5).

The number of unsatisfied clients incrementing their has set from the transmission of a symbol coded with piggyback symbol equals the number of arcs which are added in $G$. Clearly adding more arcs in $G$ can potentially increase the number of edges being added in $K$ after each iteration, and minimize the number of partitions in the heuristic minimum clique partition solution on graph $K$.

\subsection{UCIC Computational Complexity}
Step 3 of the algorithm is the main contribution of UCIC. We first analyze the computation complexity of UCIC for a worst case scenario. Consider an index coding problem where $|\beta|=n$. For such problem the order of magnitude for the computational cost of step 3 is given as $n^3$. 

The while loop will run at least $n-1$ iteration. After each iteration of while loop, at least one vertex is removed from graph $G$. At each iteration of the while loop, the graph is getting smaller, therefore the order of magnitude for total number of computation steps is bounded by the sum of cubed natural numbers and given as $\frac{n^4}{4}$, which is expressed as $\mathcal {O}(n^4)$ using the big O notation, assuming computational complexity for step 1 is bounded as $\mathcal {O}(n^3)$ . Let's consider a best case scenario when $\ell^*=1$, i.e. $K$ is a complete graph. For such class of problems no piggyback symbol can be found, and the computational complexity will be given as $\mathcal {O}(n)$. 

Due to the diversity of various classes of index coding problem, the average computation cost of UCIC is infeasible to evaluate. Nonetheless, the worst case computational complexity of $\mathcal {O}(n^4)$ for UCIC is of similar order as that of other algorithms used to solve the index coding problem. LDG has an average computational complexity of $\mathcal {O}(n^3)$~\cite{Birk06}, while the worst case computational complexity of sequential coloring algorithm is given as $\mathcal {O}(n^4)$~\cite{Dong11}.

UCIC constructs code over the finite field $\mathsf{F}_2$, which requires the simple XOR addition for encoding and decoding, and therefore save the clients from the high decoding cost of using Gaussian elimination associated with decoding codes over finite field size given as $\mathsf{F}_{q>2}$. The bipartite index coding scheme for example constructs codes over $\mathsf{F}_{q\gg2}$~\cite{Tehrani12}.

\begin{figure*}
\begin{center}
\subfigure[]{\includegraphics[width =
0.47\textwidth]{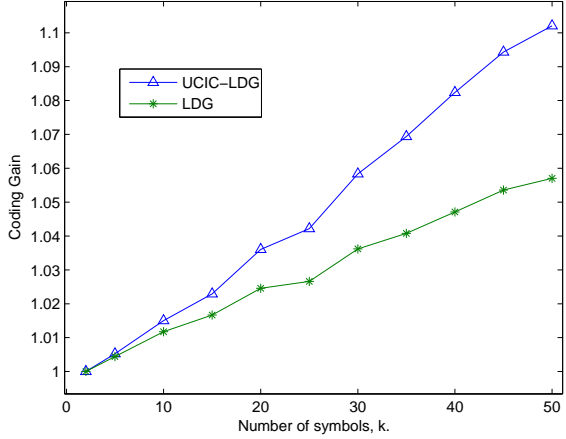}\label{fig:UCIC-LDGa}} 
\subfigure[]{\includegraphics[width =
0.47\textwidth]{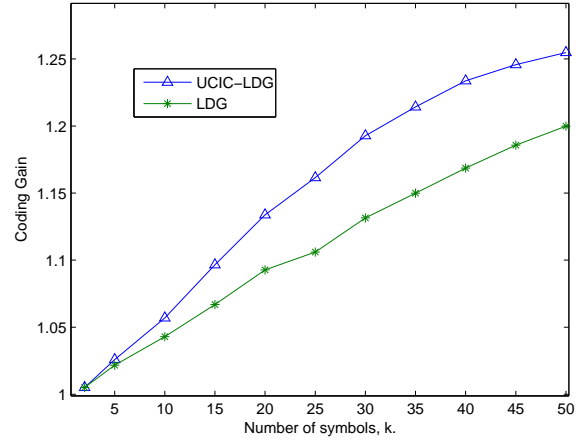}\label{fig:UCIC-LDGb}} 
\end{center}
\caption{Coding gain comparison of UCIC-LDG and LDG algorithms for (a) $p_{has}=0.05$ and (b) $p_{has}=0.1$.}
\label{fig:UCIC-LDG}
\end{figure*}

\begin{figure*}
\begin{center}
\subfigure[]{\includegraphics[width =
0.47\textwidth]{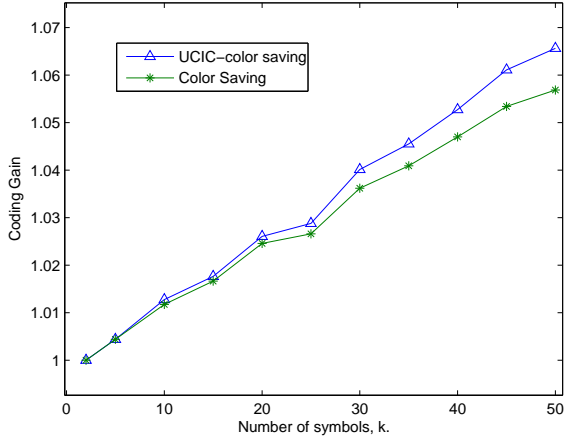}} 
\subfigure[]{\includegraphics[width =
0.47\textwidth]{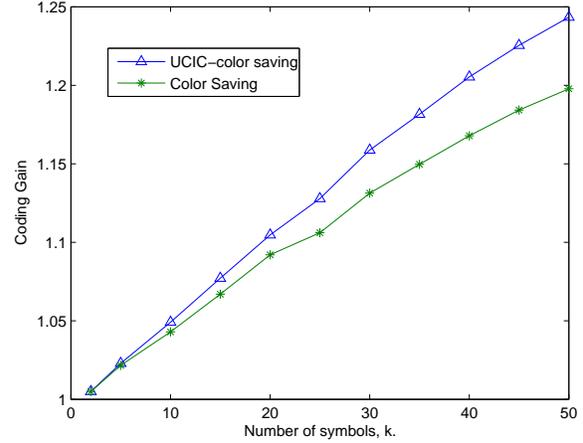}} 
\end{center}
\caption{Coding gain comparison of UCIC-color saving and color saving algorithms for (a) $p_{has}=0.05$ and (b) $p_{has}=0.1$.}
\label{fig:UCIC-CS}
\end{figure*}

\section{Simulation}\label{section:simulation}
\newtheorem{proposition}{Proposition}
We verify the higher coding gain of UCIC with LDG~\cite{Birk98, Birk06} and color-saving~\cite{Chaudhry08, Hassin94} algorithms by comparing it with UCIC-LDG and UCIC-color saving algorithms. For UCIC-LDG, at step 1 of the algorithm given in Table~\ref{table:ucic_pseudocode}, we run the LDG algorithm, similarly for UCIC-color saving we run the color saving algorithm in step 1. We construct simulators to verify the performance gain of UCIC.

The coding gain improvement of UCIC-LDG over LDG is shown in Figure~\ref{fig:UCIC-LDG}. The results show that UCIC-LDG outperforms LDG and that such performance gain of UCIC-LDG is more evident when the graph $G$ is sparse. We observe similar performance gain of UCIC-color saving over color saving, as shown in Figure~\ref{fig:UCIC-CS}. However in this figure we observe that UCIC-color saving has higher coding gain when the graph $G$ is dense. In both cases, UCIC outperforms LDG and color saving minimum clique partition algorithms, and its coding gain is dependent on the algorithm used in step 1 of the UCIC algorithm.

\begin{proposition}
UCIC coding gain is always equal or higher compared to the coding gain of any heuristic/ approximation minimum clique partition algorithm on graph $K$ used to solve the index coding problem.
\end{proposition}
\begin{proof}
If UCIC is unable to find a piggyback symbol, it then resorts to the solution constructed by the heuristic/ approximation minimum clique partition algorithm in step 5 of the UCIC algorithm. For such instance, UCIC coding gain will be equal to the coding gain of the heuristic/ approximation minimum clique partition algorithm. 

If UCIC is able to find piggyback symbols, then as shown in the simulation results, it has the potential to outperforms heuristic/ approximation minimum clique partition algorithms on graph $K$. This completes the proof.
\end{proof}

\begin{table}
\begin{center}
\caption{Characterizations of Graphs $K$ and Digraphs $G$ with Near-Extreme $\ell^*$.}
\label{table:extreme_rate}
\begin{tabular}{|l|p{4cm}|p{3cm}|}
\hline
$\ell^*$ & Graph $K$ & Digraph $G$ \\ \hline
1				 & $K$ is a complete graph. & $G$ is a complete graph.\\ \hline
2				 & $K$ is not a complete graph and $\overline{K}$ is 2-colorable. & $G$ is not a complete graph and $\overline{G}$ is fairly 3-colorable. \\ \hline
$n-2$		 & $K$ has a maximum matching of size two, does not contain graph $F$ (Figure~\ref{fig:graph_F}) as a subgraph and $k\geq 6$. & Unknown. \\ \hline
$n-1$		 & $K$ is a star graph. & Unknown. \\ \hline
$n$ 		 & $K$ has no edges. & $G$ has no cycles. \\
\hline
\end{tabular}
\end{center}
\end{table}

\begin{figure}
\begin{center}
\includegraphics[width = 0.35\textwidth]{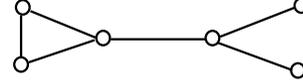}
\end{center}
\caption{The forbidden subgraph $F$.} \label{fig:graph_F} 
\end{figure}

\section{Constructing $\mathcal{D^*}$ using UCIC} \label{section:optimal}

In this section we demonstrate that UCIC can construct $\mathcal{D^*}$ for all classes of unicast data flow index coding problems where $\mathcal{D^*}$ is known. Index code $\mathcal{D^*}$ can be constructed for single-uniprior side information graph~\cite{Ong12} and near-extreme rates~\cite{Dau14}. 

It should be clarified here that in the work of Berliner and Langberg~\cite{Berliner11}, the authors find the value of $\mathsf{minrk_2}(G)=\ell^*$ for outerplanar graph but do not propose a coding scheme to construct $\mathcal{D^*}$. Knowledge of $\ell^*$ does not necessitate solution $\mathcal{D^*}$ to be known. Therefore we do not study solution $\mathcal{D^*}$ for outerplanar graph in this section.

\subsection{Near-extreme Rates~\cite{Dau14}}
In the work of Dau \textit{et al.}, the authors characterize various undirected graph $K$ and digraph $G$ whose minimum index code length $\ell^*$ is equal to 1, 2, $k-2$, $k-1$ and $k$, as summarized in Table~\ref{table:extreme_rate}. For graph $K$, symmetric data request is assumed, i.e. if client $c_i$ has symbol $p_j$, then $c_j$ has $p_i$.

The NP-hardness of minimum clique partition applies when $\varphi(G)\geq 3$. For the graphs $G$ and $K$, $\ell^*=1$ when the graph is a complete graph, a heuristic minimum clique partition algorithm can find an optimal solution. Similarly a heuristic minimum clique partition algorithm can find an optimal solution when $\ell^*=2$ for graph $G$. Constructing $\mathcal{D^*}$ when $\ell^*=2$ for digraph $G$ is a NP-complete problem~\cite[Theorem 5.2]{Dau14}. 

When $K$ has no edges, then this implies that $H_i=\emptyset$, $\forall H_i$, and the index code solution of $\mathcal{D^*}=\mathcal{P}$ is trivial. When $K$ is a star graph, a heuristic minimum clique partition algorithm will partition the graph in to $n-1$ cliques, of which exactly one clique will be of size equal to two vertices with the rest of the partitioned cliques of size equal to one. Therefore step 1 and 5 of the UCIC scheme will construct index code $\mathcal{D^*}$ of length $\ell^*=n-1$.

For graph $K$ with maximum matching $\delta(K)=2$ and $k\geq 6$, a heuristic minimum clique partition algorithm will partition $K$ in to two cliques of size equal to two vertices, with the rest of partitioned cliques of size equal to one. One can use proof by contradiction to prove that a heuristic minimum clique partition algorithm will not partition the graph in to more than or equal to three disjoint cliques of size equal to two because if this was the case then this implies that $\delta(K)\neq 2$, leading to contradiction. Such a graph $K$ with $\delta(K)=2$ can not be partitioned with a clique of size equal to three, because subgraph $F$ is a forbidden. It is easy to verify that it is not possible to have a partitioned clique in $K$ of size larger than or equal to four under the constraint of $\delta(K)=2$ and $k\geq 6$.

\subsection{Single-uniprior Side Information~\cite{Ong12}}
In a single-uniprior side information the has set is characterized as $H_i\cap H_j=\emptyset, \forall i\neq j$ and $|H_i|=1, \forall i$. 

\begin{lemma}\label{lemma:uniprior_unicast}
For the single-uniprior single-unicast transmission index coding problem represented by an information flow graph $I$, there can exist a maximum of one strongly connected component (SCC) in any edge disjoint subgraph. Further such SCC is also a chordless cycle and a non-SCC disjoint subgraph cannot exist in $I$.
\end{lemma}
\begin{proof}
For any vertex $v_i$ in $I$ for single-uniprior single-unicast transmission index coding problem, there exist only one outgoing arc, $d_{out}(i)=1$, for vertex $v_i$, and only one incoming arc, $d_{in}(i)=1, \forall i$. As every client in $I_u$ has a singleton want set, and single-uniprior side information, i.e. singleton unique has set. 

We use proof by contradiction to prove that every SCC is a chordless cycle in $I$. Let's consider a SCC which has one edge disjoint subgraph in $I$ and assume that this disjoint SCC in $I$ is not a chordless cycle. Then clearly this SCC should have at least one of the vertex with more than one outgoing edge, but all vertices in $I$ have one outgoing edge, therefore a disjoint SCC in $I$ is always a chordless cycle.

Similarly using the property of $d_{out}(i)=1$ and $d_{in}(i)=1$ it can be shown using proof by contradiction that a maximum of one SCC can exist in a disjoint subgraph, and a non-SCC disjoint subgraph cannot exist in $I$.
\end{proof}

\begin{theorem}\label{theorem0}
UCIC constructs $\mathcal{D^*}$ for single-unicast and unicast transmission index coding problem with single-uniprior side information.
\end{theorem}
\begin{proof}
Based on Lemma~\ref{lemma:uniprior_unicast} let's consider one such information flow digraph $I_u$ for single-unicast transmission with single-uniprior side information. First let's assume that $I_u$ has only one chordless cycle (SCC). The corresponding IDC graph $K$ for such a problem would have $k$ cliques of size equal to one vertex, and for every vertex $v_j$ corresponding to a symbol $p_j$, there always exists a piggyback symbol $p_g$ given as $H_j=\{p_g\}$. Transmission of $p_j\oplus p_g$ will satisfy the request of $c_j$, and increment the cache of $c_i$, $c_i : H_i=\{p_j\}$. This arises from the property of single-uniprior side information, as exactly one client (which we call $c_i$) has $p_j$ but not $p_g$. Continuing this way the code generated by UCIC will be given as $\mathcal{D^*}=\{p_j\oplus p_g, p_g\oplus p_i, p_i\oplus p_u, \ldots\}$ of length $|\mathcal{D^*}|=k-1$~\cite[Theorem 2]{Ong12}.

Now let's consider $I$ with $\xi$ edge disjoint SCCs. We had shown that UCIC is optimal for $I$ with one SCC, we now need to show that for $I$ with $\xi$ disjoint SCCs, UCIC will save exactly $\xi$ transmissions. Since it is a property of the UCIC that it should attempt to satisfy the want set of at least one client from any given transmission, two symbols corresponding to vertices in disjoint SCCs will not be coded, as the transmission of such coded symbol will not satisfy the want set of either clients. Consecutively every symbol is coded with a symbol within the same SCC. But, as we had shown earlier for any disjoint SCC, $|\mathcal{D^*}|=k-1$. Therefore UCIC will save $\xi$ transmissions from each of the $\xi$ disjoint SCCs, which correspond to the optimal index codes~\cite[Theorem 3]{Ong12} for single-unicast transmission with single-uniprior side information.

Based on Lemma~\ref{lemma:one_pk}, a unicast index coding problem can be reduced to a single-unicast index coding problem, therefore UCIC is also optimal for unicast transmission with single-uniprior side information. This completes the proof.
\end{proof}

\section{Future Work}\label{section:future}
For our future work, we would like to expand the work of UCIC for single-multicast and multiple-multicast transmission index coding problem. For the multicast transmission, characterised as $H_i\cup W_i=\mathcal{P}, \forall i$, optimal index codes known as maximum distance separable (MDS) codes already exist~\cite{Plank05}.

The complementary index coding (CIC) is designed to solve index coding problem for unicast and multiple-multicast transmission and it is centered on the following idea.

\begin{lemma}\label{lemma:cic}
For a side information graph with $\alpha$ vertex disjoint cycles, then it is possible to save at least $\alpha$ transmission, and the optimal index code length is bounded as $\ell^*\leq k-\alpha$.~\cite{Chaudhry11}
\end{lemma}

\noindent The problem of maximizing vertex disjoint cycles in $G$ is NP-hard, and therefore CIC relies on heuristic algorithm to construct $\mathcal{D}$ given as $|\mathcal{D}|=k-\alpha$. 

The bipartite index coding problem is an extension of CIC designed for single-unicast and single-multicast, and minimizes the number of transmissions by solving the PaMul optimization problem (a NP-hard problem). Therefore the number of transmissions which can be saved in a valid decomposed graph is given by the minimum outdegree of a client in a decomposed graph. We refer interested readers to~\cite[Section II]{Tehrani12} for formal definition of decomposed graphs and PaMul optimization problem.

\begin{figure}
\begin{center}
\includegraphics[width = 0.45\textwidth]{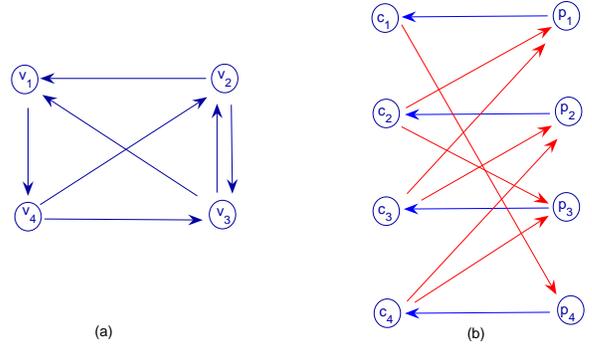}
\end{center}
\caption{The (a) side information graph $G$, and (b) bipartite graph $B$, for the index coding problem example given in Section~\ref{section:future}.} \label{fig:cic} 
\end{figure}

We use a motivating example to show that UCIC has potential performance advantage over CIC and bipartite index codes. Consider an index coding problem where $W_i=\{p_i\}$, $H_1=\{p_4\}$, $H_2=\{p_1, p_3\}$, $H_3=\{p_1, p_2\} $ and $H_4=\{p_2, p_3\}$ as shown in Figure~\ref{fig:cic}. It is easy to verify that there only exist one vertex disjoint cycle in Figure~\ref{fig:cic}(a). Similarly the minimum outdegree of the valid decomposed graph $B$ in Figure~\ref{fig:cic}(b) is given as one. Therefore CIC and bipartite index code will construct index code of length given as $\ell=3$. However for the transmission $p_1\oplus p_2\oplus p_3$, the piggyback symbol $p_1$ can be used to update the cache of $c_4$, and satisfy $c_2$ and $c_3$. Therefore UCIC constructs $\mathcal{D^*}=\{p_1\oplus p_2\oplus p_3, p_1\oplus p4\}$, with $\ell^*=2$, demonstrating its performance improvement potential over CIC and bipartite index codes.

\section{Conclusion} \label{section:conclusion}
In this paper we demonstrated index coding scheme which adds piggyback symbol in the coded symbol so that some of the unsatisfied clients can update their cache, which we call updated clique index coding (UCIC). Through a motivating example to present intuition of UCIC and simulation results we showed that UCIC has higher coding gain over existing index coding schemes for unicast data flow. 

We further showed that UCIC construct minimum length index code for those classes of index coding problems where minimum length index codes are known. UCIC uses the simple concept of coding piggybacking symbol using XOR addition and does not have encoding, decoding and algorithmic computational overhead. In our list of future work we showed that UCIC can be extended for index coding problem with single-multicast and single-multicast data flow and has potential of higher coding gain over bipartite index coding and complementary index coding schemes.
\bibliographystyle{IEEEtranS}
\bibliography{IEEEabrv,mainJ}

\appendix
\subsection{Graph Terminologies}
Any graph $D$ is represented by the set of its vertices and edges, which may or may not be directed edges, and is given as $D=(V,E)$, where $V=\{v_1,\ldots,v_m\}$, and $E=\{e_1,\ldots,e_\kappa\}$.

\begin{itemize}

\item A \textit{path} from $v_1$ to $v_\eta$ is a sequence $P=v_1,e_1,v_2,e_2,\ldots,e_{\eta-1},v_\eta$ of alternating vertices and edges such that for $1\leq \mu < \eta$, $e_\mu$ is incident with $v_\mu$ and $v_{\mu+1}$. If $v_1=v_\eta$, then $P$ is said to be a \textit{cycle}. A cycle given such that there exists only one path from any pair of vertices in the cycle is known as \textit{chordless cycle}.

\item Two vertices $v_1$ and $v_2$ are said to be \textit{strongly connected} if there is a directed path from $v_1$ to $v_2$ and from $v_2$ to $v_1$. The \textit{strongly connected component} (SCC) of a graph are the maximal disjoint subset of vertices such that that each subset of vertices is strongly connected. A graph cycle is a subset of SCC.

\item The \textit{complementary graph} $\overline{D}=(V,\overline{E})$ of $D$, has the same vertices as $D$, and edges between every pair of vertices except those pairs of vertices for which an edge exist in the graph $D$.

\item A \textit{clique} $C$ (also known as \textit{complete graph}) is a subgraph of $D$, $C\subseteq D$, such that there exist undirected edge between all pairs of vertices in $C$.

\item The \textit{degree} of a vertex $v_i$ is the number of edges incident with $v_i$. The number of outgoing and incoming arcs of $v_i$ are called the \textit{outgoing degree} $d_{out}(i)$ and \textit{incoming degree} $d_{in}(i)$ of $v_i$ respectively.

\item An \textit{independent edge set} is a set of edges such that no pair of edges share a common vertex.

\end{itemize}

\subsection{Graph Optimization Problems}
\begin{itemize}

\item \textit{Maximum independent (edge) set problem, $\delta(D)$.} Given the graph $D$ find an independent set of maximum cardinality. Also known as \textit{maximum matching}. 

\item \textit{Clique Number, $\omega(D)$.} A clique with maximum number of vertices in $D$. The $\delta(D)$ and $\omega(\overline{D})$ problems are equivalent.

\item \textit{Minimum clique partition problem, $\varphi(D)$.} Given the graph $D$, partition $V$ into disjoint subsets $V_1, V_2,\ldots, V_k$ such that, for $1\leq i\leq k$, the subgraph induced by $V_i$ is a complete graph, and $k$ is minimized. 

\item \textit{Minimum vertex coloring problem, $\chi(D)$.} Given the graph $D$, assign $k$ colors to all vertices of $D$ such that no two adjacent vertices should have the same color, and $k$ is minimized. The solution of this problem is known as the chromatic number of $D$. The $\chi(D)$ and $\varphi(\overline{D})$ problems are equivalent.

\end{itemize}

\end{document}